\documentclass[acmsmall,sigconf]{acmart}
\usepackage{lipsum}

\AtBeginDocument{%
  \providecommand\BibTeX{{%
    \normalfont B\kern-0.5em{\scshape i\kern-0.25em b}\kern-0.8em\TeX}}}

\newcommand{\myname} {{\em DRLC}}

\copyrightyear{2021}
\acmYear{2021}
\setcopyright{acmcopyright}\acmConference[SIGIR '21]{Proceedings of the 44th International ACM SIGIR Conference on Research and Development in Information Retrieval}{July 11--15, 2021}{Virtual Event, Canada}
\acmBooktitle{Proceedings of the 44th International ACM SIGIR Conference on Research and Development in Information Retrieval (SIGIR '21), July 11--15, 2021, Virtual Event, Canada}
\acmPrice{15.00}
\acmDOI{10.1145/3404835.3463228}
\acmISBN{978-1-4503-8037-9/21/07}
\begin{document}

\title{De-Biased Modelling of Search Click Behavior with Reinforcement Learning}

\author{Jianghong Zhou}

\affiliation{%
  \institution{Emory University}
}
\email{jianghong.zhou@emory.edu}
\author{Sayyed M. Zahiri}
\affiliation{%
  \institution{The Home Depot}
}
\email{mzahiri@gatech.edu}
\author{Simon	Hughes}
\affiliation{%
  \institution{The Home Depot}
}
\email{simon_hughes@homedepot.com}

\author{Khalifeh Al Jadda}
\affiliation{%
  \institution{The Home Depot}}
\email{KHALIFEH_AL_JADDA@homedepot.com}

\author{Surya Kallumadi}
\affiliation{%
  \institution{The Home Depot}}
\email{surya@ksu.edu}

\author{Eugene Agichtein}
\affiliation{%
  \institution{Emory University}
  }
\email{eugene.agichtein@emory.edu}



\begin{abstract}
Users’ clicks on Web search results are one of the key signals for evaluating and improving web search quality, and have been widely used as part of current state-of-the-art Learning-To-Rank(LTR) models. With a large volume of search logs available for major search engines, effective models of searcher click behavior have emerged to evaluate and train LTR models. However, when modeling the users’ click behavior, considering the bias of the behavior is imperative. In particular, when a search result is not clicked, it is not necessarily chosen as not relevant by the user, but instead could have been simply missed, especially for lower-ranked results. These kinds of biases in the click log data can be incorporated into the click models, propagating the errors to the resulting LTR ranking models or evaluation metrics. In this paper, we propose the De-biased Reinforcement Learning Click model (\myname{}). The \myname{} model relaxes previously made assumptions about the users’ examination behavior and resulting latent states. To implement the \myname{} model, convolutional neural networks are used as the value networks for reinforcement learning, trained to learn a policy to reduce bias in the click logs. To demonstrate the effectiveness of the \myname{} model, we first compare performance with the previous state-of-art approaches using established click prediction metrics, including log-likelihood and perplexity. We further show that \myname{} also leads to improvements in ranking performance. Our experiments demonstrate the effectiveness of the \myname{} model in learning to reduce bias in click logs, leading to improved modeling performance and showing the potential for using \myname{} for improving Web search quality.
\end{abstract}


\keywords{web search click models, reinforcement learning for click modeling, de-biased estimation of search click behavior}


\maketitle

\section{Introduction}
To improve web search, it is important to understand how the
users interact with web search engines.  Clicking on search results is one of the most important interaction data, which forms the basis of many important user behaviors, like reformulating or switching queries, clicking on different items, and browsing the search results. One powerful way to utilize these click logs is to construct a click model to measure and predict clicks on existing or future results\cite{borisov2016neural}. A click model can predict future clicks of other users, help train a learning to rank (LTR) model, and allow to automatically evaluate search result quality. However, modeling users' clicks is a challenging task, because the click logs are observational data, collected in-situ with a live search engine, and exhibits multiple biases\cite{joachims2017unbiased}. Previous research on click modeling and prediction did not directly address this issue or addressed the biases using heuristics, resulting in poor model performance on live (unseen) query traffic\cite{joachims2017unbiased}.  

As a summary, previous research focused on two directions. The first direction is the Probabilistic Graphical Model Framework Based Methods (PGM) \cite{koller2009probabilistic}.  Those methods model the search process as a sequence of events, predicting the click based on some probability models and assumptions. They are flexible and interpreted. The second direction considers a different representation of the events \cite{borisov2016neural}. The searching process is represented by some vectors. This form allows the users to consider a variety of features easily and to feed them to some rather stronger learning models, such as neural networks \cite{chakraborty2000bishop}. However, the drawbacks of these two directions are obvious. The PGM methods are limited by a rather weak learning model and fewer features. The second direction cannot consider the bias issue in an interpreted way.

To overcome the aforementioned issues, we propose our new model, \myname{} for training unbiased (or less biased) click models. \myname{} is also a PGM-based method. Therefore, \myname{} can be organized in a flexible way for different ranking scenarios and generate an interpretive model to reduce a variety of biases. However, in comparison to the previous PGM method, \myname{} is constructed by a more dynamic system, which is the reinforcement learning \cite{sutton2018reinforcement,zhou2020rlirank,zhou2020diversifying}. This allows \myname{} to takes advantage of stronger learning models (neural networks). 

Another contribution of this paper is we propose a posteriori method to learn the observation bias. Previous approaches normally apply some priori probability models to estimate the observation probability or reduce the bias \cite{dupret2008user,agarwal2019estimating}. Those methods highly rely on the selected priori models, which is hard to be general. In this paper, we concentrate on the posteriori knowledge. When users are browsing the SERP (search engine result page), the latter part is naturally unobserved. Therefore, if we use an observation window to augment the dataset, we can have some data with more bias and some data with less bias. This separation is helpful to capture the bias. \myname{} uses this method to reduce the bias.

\section{Related Work}

We discuss two important methods related to our work: PGM framework based methods and neural click model.\\

\noindent\textbf{Probabilistic Graphical Models:}
Most of the traditional click models are based on the PGM framework \cite{koller2009probabilistic}. In PGM, the users' interactions are organized as a sequence of events, such as document examinations, skips, and clicks. The most frequent events that those models consider are document examinations and clicks. Since it is instinctively correct that a user's clicks should be led by a document examination, these two events are highly interdependent. Based on this assumption, some important click models, including CCM, DCM, DBN, UBM  are proposed \cite{dupret2008user,guo2009click,chapelle2009expected}.\\

\noindent\textbf{Neural Click Models:}
The neural click model (NCM) is a method based on artificial neural networks. The input data of the artificial neural network is generally one or many vectors, even high dimensional tensor. Therefore, 
different from the PGM based method, NCM represents the clicking
process with vectors \cite{borisov2016neural}. \cite{borisov2016neural} attempts to use many different neural networks to construct the click models, including Long Short-Term Memory (LSTM) \cite{huang2015bidirectional} and recurrent neural networks (RNN) \cite{yang2016recurrent}. Their empirical evaluations showed that the LSTM version has better performance.

Additionally, NCM also expands to some complicated click models. For example, \cite{chen2020context} develops a session search click model (CACM), which leverages the signals from previous search iterations to predict the next search iteration of the search session.

\section{The Proposed Method}
In this part, we introduce the proposed \myname{} model. We first introduce the implementation of the \myname{}'s components, namely the value networks, which include a bias network, and a de-biased network. Then, we overview the proposed reinforcement learning (RL) approach to training a de-biased click model.


\subsection{Value Network Implementation}
We chose to use Convolutional Neural Networks (CNNs) as our value networks. The first CNN $C_1$is a bias network, which means this network considers the observation bias: Some documents are not clicked because they are not observed. The input features are bias features $B$ and the document features $D$. $B$ is a vector with $1$ or $0$. $B$ represents the observation situation of the search results. If the item is observed, the value is $1$ or it is $0$. We use an observation window to observe the SERP. We first assume the documents browsed by the window are observed. Then, by comparing the predictions from two value networks, we further update $B$. $D$ is a vector, representing the features of a document. Those features are related to both the query and the document, such as the frequency of the query appearing in the document. The output of $C_1$ is the click estimation with bias or the possibility of clicking this document. The second CNN $C_2$ is a de-biased network, whose output is the de-biased click estimation or the possibility of clicking the item under the de-biased setting. The input of $C_2$ is just $D$.
`
The structure of the value networks is summarized in Fig. 1(a).
\begin{figure*}[htb]
\centering
\[
\begin{array}{cc}
    \includegraphics[width=240pt]{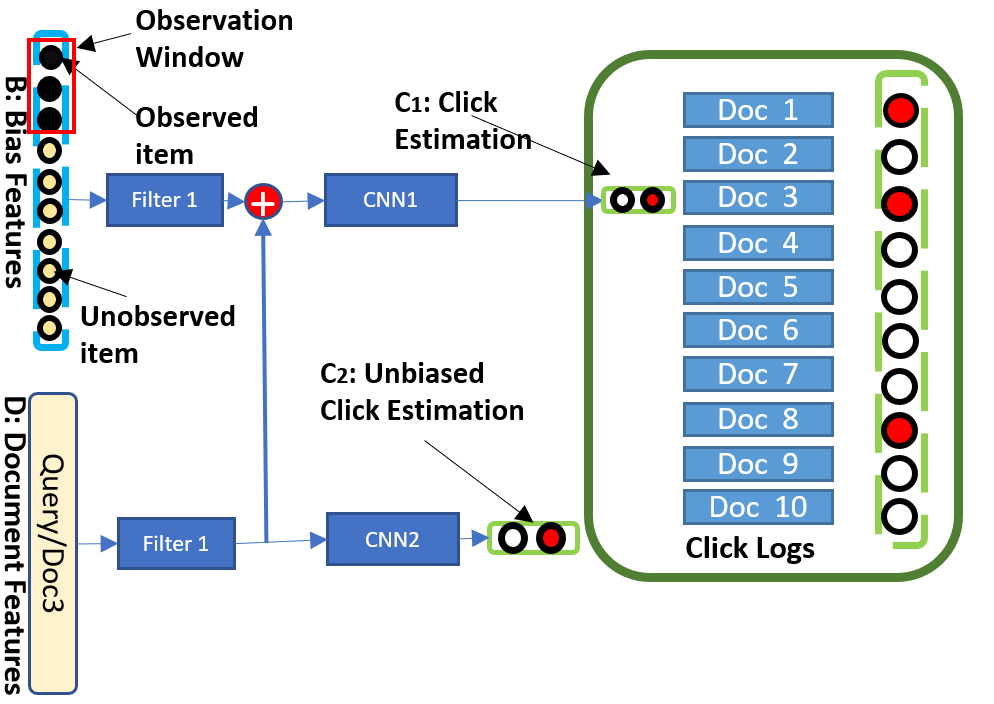}
  &
 \includegraphics[width=240pt]{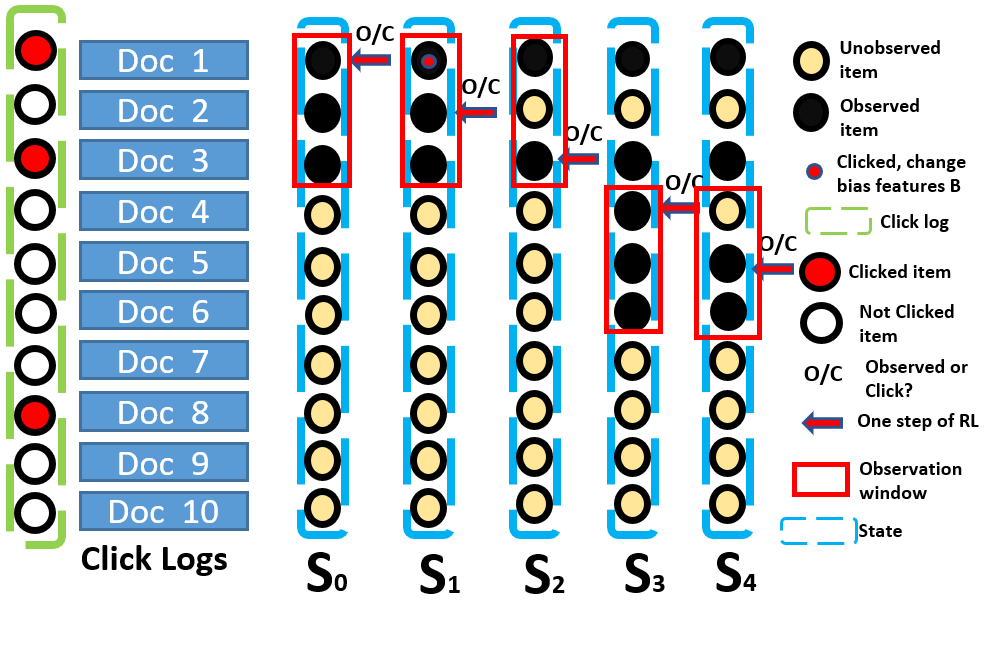}
  \\
  (a) & (b) 
\end{array}
\]
\vspace{-3mm}
\caption{(a) The structure of the \myname{} value networks.; (b) Illustration of the \myname{} Reinforcement Learning process.}
\label{fig:rli-rank-results}
\end{figure*}

\noindent\textbf{Pre-training: }Since two networks focus on two aspects of the dataset, we pre-train them in different ways. We pre-train $C_1$  with the whole training dataset, which is a highly biased dataset. For the observation features $B$, we assume that the users observe the documents sequentially with an observation window. If the documents appeared in the window before, they are denoted as observed. If not, they are not observed. Only when all the documents in the observation window are examined, the window moves to the next position, the process is presented in Fig. 1(b). The size of the observation window is decided by the empirical estimation (In this paper, the size of the window is 3.). $C_2$ is trained by a de-biased dataset. We assume the documents before the last clicked document is more likely to be observed. Therefore, we only pre-train $C_2$ with the documents before the last clicked document.

\noindent\textbf{CNN architecture}
 The input of the $C_1$ is an $100\times 1$ vector (Bias features $B$) and a $56\times 1$ vector (Document features $D$). The bias features are demonstrated in \textbf{Pre-train} part. The document features are generated from the URL provided in the dataset. The features we selected are identical to the ones listed in \cite{qin2010letor}. Firstly, we apply two filter layers to join the features from $D$ and $B$. For each CNNs, they have three convolutional blocks. Each convolutional block contains 16 filters of kernel $3\times1$ with stride 1, a batch normalization layer, and a ReLU layer. The output layer is a fully connected network. The loss function is a softmax function. The structures of CNNs cannot be too deep, considering the scalability of the click model.

After pre-training, the networks are initialized to be trained by the RL method. 
\subsection{Reinforcement Learning (RL) Framework}
The RL framework of \myname{} is illustrated in Fig. 1(b). The components of the RL are  defined as follows:

\noindent\textbf{State $s$} is the click and observation state of the documents. We assume a user's attention can be modeled by an observation window, where users observe items in the observation window with a higher possibility. 

\noindent\textbf{Action $a$} is selecting the state of a document at the position. Those states include: (Observed, Click), (Observed, Not click), and  (Unobserved, Not click).

\noindent\textbf{Transition $T$} changes the document state in $S$ based on $a$.

\noindent\textbf{Reward $R$} is how well the estimated click probabilities match the empirical distribution observed in the click logs.
\begin{equation}
R_t =  (C_t-C_{t,1})^2+\beta O_t(C_t-C_{t,2})^2
\end{equation}
Where $t$ is the position of the document. $C_t$ is the click of the click logs. $C_{t,1}$ is the click prediction from $C_1$. $C_{t,2}$ is the click prediction from $C_2$. $\beta$ weighs the importance of the de-biased prediction The value is  set to $0.7$ based on the validation part of the experiment. $O_t$ shows that whether this document is observed. We have two assumptions of the observation. First, if the document is clicked in the click logs, it is observed. Second, if $C_{t,1}/C_{t,2}<\theta$, the document is unobserved. The first assumption is understandable. The second assumption is based on the bias effect, which is $P(O_t)C_{t,2}=C_{t,1}$. If $P(O_t)$ is small, it means the possibility of the observation is low. In this paper, we empirically set $\theta$ as 0.3.

The goal of the RL is to learn a policy $\pi^*$ to maximize $\mathbb{R}=\sum \gamma^t R_t$. In turn that means learning the {\em value} of each state, corresponding to click probability. 

\noindent\textbf{Update:} The $C_1$ are further trained by the results of the final state $S_T$, where $T$ is the total number of the documents in the click log. The $C_2$ is updated by the observed documents.
\section{Experiments and Discussion}
In this section, we introduce the datasets, metrics, and baselines.

\textbf{Datasets:} We demonstrate the effectiveness of the proposed model in two open-sourced dataset and one private dataset: ORCAS dataset \cite{craswell2020orcas}, Yandex click dataset \cite{serdyukov2013wscd2013} and the real interactive dataset from a large e-commerce website (\url{https://www.homedepot.com/}). ORCAS is a click-based dataset associated with the TREC Deep Learning Track. It covers 1.4 million of the TREC DL documents, providing 18 million connections to 10 million distinct queries. The Yandex click dataset comes from the Yandex search engine, containing more than
30 million search sessions. Each session contains at least
one search query together with 10 ranked items. The private Interactive Dataset (RID) is a 3-month search log. In this dataset, the users normally search for several queries. For each query, the search engine returns a list of products and then the user can interact with the results by clicking, adding the items to the cart and ordering. Table \ref{tab3} shows a sample of the data. Additionally, with the products' ID, we can further find the page of the products and extract the text features. The features list can be found in \cite{qin2013introducing}. 


\begin{table*}
\center
\caption{The sample of the Home Depot website's Interactive data.}\label{tab3}
\begin{tabular}{l|l|l|l|l|l|l|l|l}
\hline
visitor ID &   session id & date&time&searchterm&	click sku&atc sku&order sku&product impression	\\
\hline
1000 &  1000-mobile-1 & 6/1/2020&6:30 pm&everbilt dropcloth&2034& & & 3072|2034|2037|2036\\
1000 &  1000-mobile-1 & 6/1/2020&6:34 pm&pull down shades&3022&3022&3022&3022|2051|3042|2071\\
1001 & 1001-mobile-1 & 6/1/2020&6:36pm&	fence panel&&&&2030|1003|2029|1000\\
1001 & 1001-mobile-2 &6/1/2020&6:36pm&fince dog ears&2053&&&2055|2034|3034|2053\\
\hline
\end{tabular}
\end{table*}
\textbf{Metrics}
We evaluate the model from two aspects. The first aspect is based on the click prediction. The second aspect is based on relevance. In terms of the click prediction, we use Log-likelihood and perplexity as the evaluation methodology. In terms of the relevance analysis, if the item is clicked, its relevance score is 1, or it is 0. Then, we consider the scores calculated to predict the click as the relevance score. We rank these scores and calculate the NDCG as our relevance prediction metric \cite{borisov2016neural}.

\textbf{Baselines}
We use DBN, DCM, CCM, UBM and NCM as our baselines \cite{chapelle2009dynamic,chuklin2013evaluating,dupret2008user,guo2009efficient,borisov2016neural}. Those methods are the state-of-art click models based on PGM and neural networks. The parameter settings are based on the Pyclick package (\url{https://github.com/markovi/PyClick}). (We did not use CACM as our baseline, because it is a session search click model, not a single search click model \cite{chen2020context}. ) 

\begin{table*}
\center
\caption{The experiment results of DBN, DCM, CCM, UBM, NCM, URCM on ORCAS dataset, Yandex click dataset and the \url{https://www.homedepot.com/} e-commerce website interactive dataset (RID). The best  performance results are highlighted in bold font. All improvements are significant with $p<0.05$.}\label{tab1}
\begin{tabular}{l|l|l|l|l|l|l|l}
\hline
Dataset&Model &  Perplexity & Log-likelihood & NDCG@1&NDCG@3&NDCG@5&NDCG@10\\
\hline
{ORCAS Dataset}&DBN &  1.4628 & -0.2273&0.596&0.606&0.623&0.655\\~&
DCM &  1.4647 & -0.2894&0.609&0.618&0.639&0.662\\~&
CCM & 1.4664 & -0.2778&0.615&\textbf{0.626}&0.637&0.671\\~&
UBM & 1.4593 &-0.2203&0.599&0.608&0.628&0.656\\~&
NCM & 1.4545 & -0.2186&\textbf{0.617}&0.625&0.639&0.677\\~&
\myname & \textbf{1.4326} &\textbf{-0.2037}&0.610&0.624&\textbf{0.645}&\textbf{0.686}\\
\hline

\hline
Yandex Click Dataset&DBN &  1.3562 & -0.2789&0.702&0.724&0.766&0.841\\~&
DCM &  1.3605 & -0.3594&0.729&0.744&0.775&0.845\\~&
CCM & 1.3688 & -0.3522&0.746&0.757&0.779&0.848\\~&
UBM & 1.3422 &-0.2667&0.729&0.739&0.769&0.841\\~&
NCM & 1.3406 & -0.2522&\textbf{0.756}&\textbf{0.763}&\textbf{0.788}&0.846\\~&
\myname & \textbf{1.3283} &\textbf{-0.2393}&0.729&0.754&0.776&\textbf{0.848}\\
\hline

\hline
{RID dataset}&DBN &  1.3777 & -0.2267&0.543&0.578&0.598&0.605\\&
DCM &  1.3764 & -0.2873&0.566&0.587&0.603&0.611\\&
CCM & 1.3872 & -0.2983&0.511&0.601&0.608&0.621\\&
UBM & 1.3899 &-0.2637&0.538&0.612&0.618&0.632\\&
NCM & 1.3937 & -0.2433&0.556&0.617&0.623&0.638\\&
\myname & \textbf{1.3554} &\textbf{-0.2232}&\textbf{0.616}&\textbf{0.624}&\textbf{0.645}&\textbf{0.648}\\
\hline
\end{tabular}
\end{table*}
\textbf{Results and Discussion}
The results of the experiment are summarized in Table \ref{tab1}. The empirical results show that \myname{} outperforms all baselines in terms of clicking prediction by 3.4\% to 5.2\%. Based on the T-test, this improvement is substantial. For the ranking prediction, \myname{} outperforms the other baselines mostly.

 Based on the empirical results, \myname{} can predict the click, both in accuracy and relevance, better than the previous methods. We attribute the improvement of the click prediction to our assumptions of observation. In past, the unobserved data is hard to train, because it is almost impossible to manually label the data as observed documents or unobserved ones. However, in our RL framework, the users browse the results sequentially. It is reasonable to assume that the documents are unobserved when the users are still observing the top documents. In this way, we can train the networks to classify whether the documents are observed. Besides, when the de-biased prediction and the biased prediction have a huge difference, it also indicates that the document is highly possible to be unobserved. 
 
 Additionally, in the RL framework, the initial State is updated during the training process to further improve the value networks. 

\section{Conclusions}
In this paper, we propose \myname, which is a de-biased click model based on reinforcement learning. The model aggregates the advantages of the previous PGM methods and neural network methods, modeled by some novel assumptions of users' clicking and the observation bias. The empirical evaluation shows that \myname{} is the state-of-art method in terms of click prediction, indicating that the RL, CNNs, and the proposed assumptions in this paper are helpful to improve the performance of the click models.

In the future, we can further consider other methods to de-bias the dataset. For example, we can apply counterfactual learning to train the de-bias network, which is theoretically unbiased \cite{agarwal2019estimating}. \myname{} can also be a trainer of some novel LTR models, providing a de-biased dataset.

\bibliographystyle{ACM-Reference-Format}
\bibliography{sample-base}

\appendix

\end{document}